\documentclass[10pt]{article}
\usepackage{fullpage}
\usepackage{amsmath}
\usepackage{amssymb}
\usepackage{amsfonts}
\usepackage{comment}
\usepackage[dvips]{epsfig}
\usepackage{color}
\usepackage{cite}

\def\##1{\underline{#1}}
\def\=#1{\underline{\underline{#1}}}

\def\+
#1{\underline{\bf #1}}
\def\*#1{\underline{\underline{\bf #1}}}

\def\r#1{(\ref{#1})}
\def\l#1{\label{#1}}
\def\c#1{\cite{#1}}

\def\le{\left(}
\def\ri{\right)}
\def\les{\left[}
\def\ris{\right]}

\def\.{\mbox{ \tiny{$^\bullet$} }}

\def\eps{\varepsilon}

\def\epso{\eps_{\scriptscriptstyle 0}}

\def\muo{\mu_{\scriptscriptstyle 0}}

\def\ko{k_{\scriptscriptstyle 0}}

\def\ux{\hat{\#u}_x}
\def\uy{\hat{\#u}_y}
\def\uz{\hat{\#u}_z}

\def\calA{{\cal A}}
\def\calB{{\cal B}}

\begin{document}

\begin{center}

\LARGE{ {\bf  Exorcizing ghost  waves  
}}
\end{center}
\begin{center}
\vspace{10mm} \large

 Tom G. Mackay\footnote{E--mail: T.Mackay@ed.ac.uk.}\\
{\em School of Mathematics and
   Maxwell Institute for Mathematical Sciences\\
University of Edinburgh, Edinburgh EH9 3FD, UK}\\
and\\
 {\em NanoMM~---~Nanoengineered Metamaterials Group\\ Department of Engineering Science and Mechanics\\
Pennsylvania State University, University Park, PA 16802--6812,
USA}\\
 \vspace{3mm}
 Akhlesh  Lakhtakia\footnote{E--mail: akhlesh@psu.edu}\\
 {\em NanoMM~---~Nanoengineered Metamaterials Group\\ Department of Engineering Science and Mechanics\\
Pennsylvania State University, University Park, PA 16802--6812, USA}

\normalsize

\end{center}

\begin{center}
\vspace{15mm} {\bf Abstract}
\end{center}
The so-called  electromagnetic ghost waves are simply  electromagnetic
nonuniform
 plane waves, whose
association  with both propagating and evanescent fields has long been known, even for
isotropic dielectric materials that are
 non-dissipative.
\vspace{5mm}

\vspace{5mm}

\vspace{10mm}

In recent reports \c{Nar_PRA,Nar_conf,Nar_arxiv}, the notion of ``electromagnetic ghost waves" has been presented. These are claimed to represent a
new form of planewave propagation, supported by non-dissipative (and non-active) biaxial dielectric materials. The key characteristic of these waves is that the Cartesian components
of their wavevector   are complex valued, despite the permittivity  dyadic of the biaxial material being positive definite. Accordingly, these waves are associated with both propagating and evanescent fields. 
 
 In fact, the ghost waves described in
 Refs.~\citenum{Nar_PRA,Nar_conf,Nar_arxiv} are simply nonuniform plane waves. The defining characteristic of a nonuniform plane wave is that the planes of constant phase do not  coincide with  the planes of constant amplitude (as they do in the case of a uniform plane wave). 
 The existence of nonuniform plane waves is well established for both
 non-dissipative and dissipative materials, including isotropic dielectric materials, as is described in  textbooks \c{Chen,EAB}.
 
 The  wavevector $\#k$ of a nonuniform plane wave may be decomposed as
 \begin{equation}
 \#k = \#k_r + i \#k_i,
 \end{equation}
 with $\#k_r, \#k_i \in \mathbb{R}^3$ and $\#k_r \times \#k_i \neq \#0$.
Thus, $\#k$ has complex-valued components, even  for non-dissipative materials.
The corresponding electric and magnetic field phasors, written as
\begin{equation} \l{pw}
\left.
\begin{array}{l}
\#E (\#r) = \#E_{\text{o}} \exp \le i \#k_r \. \#r \ri  \exp \le - \#k_i \. \#r \ri \vspace{8pt}\\
\#H (\#r) = \#H_{\text{o}} \exp \le i \#k_r \. \#r \ri  \exp \le - \#k_i \. \#r \ri 
\end{array}
\right\}
\end{equation}
with amplitudes $\#E_{\text{o}}, \#H_{\text{o}} \in \mathbb{C}^3$, have a propagating aspect
determined by $\#k_r$ and an evanscent aspect determined by $\#k_i$.

Three straightforward examples are presented here of nonuniform planewave propagation in dielectric
materials.
The following boundary-value problem is envisaged: A half-space $z>0$ is filled with a generally anisotropic dielectric material, characterized by
the relative permittivity dyadic
 $\=\eps_\calA$. The half-space $z<0$ is  filled with an isotropic dielectric material, characterized by
the relative permittivity scalar
 $\eps_\calB$. The materials in both half-spaces are considered to be non-dissipative.
A uniform plane wave 
is launched from a distant source in half-space $z<0$, which propagates  towards the interface $z=0$.
Since the  material occupying half-space $z<0$  is non-dissipative and the plane wave is uniform, 
the Cartesian components of the wavevector in $z<0$ are real valued.
Our attention is focused on the corresponding nonuniform planewave(s) 
in the half-space $z>0$, with
wavevector(s) 
$\#k = k_x \ux + k_y \uy + k_z \uz $. By invoking standard boundary conditions \c{Chen}, it is  inferred that
$k_x, k_y \in \mathbb{R}$.
The nonuniform plane wave(s) in $z>0$  can have both propagating and evanescent aspects, as we now demonstrate. 

The source-free Maxwell curl postulates in the half-space $z>0$ yield
\begin{equation} \l{MP}
\left.
\begin{array}{l}
\#k \times  \#E_{\text{o}}  = \omega \muo  \#H_{\text{o}} \vspace{8pt}\\
\#k \times  \#H_{\text{o}}  = - \omega \epso \=\eps_\calA \. \#E_{\text{o}} 
\end{array}
\right\}, 
\end{equation}
with $\epso$ and $\muo$ being the permittivity and permeability of free space, and $\omega$ being the angular frequency.
The wavevectors $\#k$ are deduced by solving the  dispersion relation that follows from
  seeking non-trivial solutions to 
  Eqs.~\r{MP} for a specific relative permittivity dyadic  $ \=\eps_\calA$.

\begin{enumerate}
\item
 First, we consider the case of 
 nonuniform planewave propagation
in  an isotropic dielectric material characterized by the relative permittivity $\eps_\calA$, i.e., $\=\eps_\calA = \eps_\calA \=I$.
 The corresponding dispersion relation  yields \c{Chen}
 \begin{equation} \l{iso_soln}
 k_z = \sqrt{\ko^2 \eps_\calA - k^2_x - k^2_y },
 \end{equation}
 wherein $\ko = \omega \sqrt{\epso \muo}$ is the free-space wavenumber. Thus, $k_z$ is purely imaginary provided
that $k^2_x + k^2_y > \ko^2 \eps_\calA$; otherwise $k_z$ is purely real valued.
\item
Second, we consider nonuniform  planewave propagation 
in  a uniaxial dielectric material with relative permittivity dyadic \c{EAB}
\begin{equation}
\=\eps_\calA = \eps_\calA^t \=I  + \le \eps_\calA^z - \eps_\calA^t \ri \uz \, \uz,
\end{equation} 
whose optic axis is directed along the unit vector $\uz$, 
and we assume
that $\eps_\calA^t,\eps_\calA^z>0$.  The corresponding dispersion relation yields the following two solutions  \c{EAB}:
\begin{equation} \l{uniax_soln}
k_z = 
\left\{
\begin{array}{l}
  \sqrt{\ko^2 \eps_\calA ^t - k^2_x - k^2_y } \vspace{8pt}\\
\sqrt{\eps_\calA^t
\le   \displaystyle{ \ko^2 - \frac{k^2_x + k^2_y }{\eps^z_\calA} }\ri}
\end{array}
\right. .
\end{equation}
The upper solution holds for an \emph{ordinary} plane wave and is identical in character to the solution \r{iso_soln}
for an isotropic dielectric material.
For the lower solution,  which holds for an \emph{extraordinary} plane wave, 
 $k_z$ is purely imaginary provided
that $k^2_x + k^2_y > \eps^z_\calA  \ko^2  $; otherwise, $k_z$ is purely real valued.

\item Third, we consider nonuniform  planewave propagation 
for  a biaxial dielectric material with relative permittivity dyadic \c{EAB}
\begin{equation}
\=\eps_\calA = \eps^x_\calA \, \ux \, \ux + \eps^y_\calA \, \uy \, \uy   + \eps^z_\calA \, \uz \, \uz.
\end{equation} 
 The corresponding dispersion relation yields the following  four solutions  \c{EAB}:
 \begin{equation} \l{biax_soln}
  k_z =
 \left\{\begin{array}{l}
  \displaystyle{ \sqrt{ \frac{-\beta \pm \sqrt{\beta^2 - 4  \eps^z_\calA  \gamma}}{2  \eps^z_\calA }  }}
 \\[12pt]
 \displaystyle{
- \sqrt{ \frac{-\beta \pm \sqrt{\beta^2 - 4  \eps^z_\calA  \gamma}}{2  \eps^z_\calA }  }}
\end{array}\right.
\,,
 \end{equation}
 where
 \begin{equation}
 \left.
 \begin{array}{l}
 \beta = \displaystyle{ k^2_x \le \eps^x_\calA + \eps^z_\calA \ri + k^2_y \le \eps^y_\calA + \eps^z_\calA \ri - 
  \ko^2  \le \eps^x_\calA + \eps^y_\calA \ri \eps^z_\calA} \vspace{8pt} \\
  \gamma =\displaystyle{ \le k^2_x + k^2_y -   \ko^2 \eps^z_\calA \ri
   \le k^2_x \eps^x_\calA + k^2_y \eps^y_\calA 
  - \ko^2 \eps^x_\calA \eps^y_\calA  \ri  }
 \end{array}
 \right\}.
 \end{equation}
 Only  two solutions  from Eqs.~\r{biax_soln} are physically acceptable, namely those two solutions
 yielding $\mbox{Im} \les k_z \ris > 0 $. Unlike the isotropic dielectric and the uniaxial dielectric solutions represented by 
 Eqs.~\r{iso_soln}
 and \r{uniax_soln}, respectively, the biaxial dielectric solution \r{biax_soln} can deliver a complex-valued $k_z$ with non-zero real and imaginary parts. This is most easily illustrated numerically: For example, suppose that $k_x = \sqrt{2} \ko$ and
 $k_y = \ko$, which is possible if $\eps_\calB > 3$. Also, suppose that $\eps^x_\calA = 1.45$, 
$\eps^y_\calA = 1.65$,  and $\eps^z_\calA = 1.5$. Then the two physically acceptable solutions emerging from 
Eqs.~\r{biax_soln}
are $k_z = (\pm 0.032988 + 1.21151 i)/\ko$. These signify nonuniform plane waves that have both  propagating and evanescent aspects.

\end{enumerate}

 Parenthetically, characteristics  of the type often associated with certain metamaterials can arise via
  nonuniform planewave propagation. For example, a simple homogeneous isotropic dielectric material can support nonuniform planewave propagation with negative phase velocity (and orthogonal phase velocity), without the need for a complex micro- (or nano-) structure of the type often associated with certain metamaterials
\c{ML_PRB}. A richer   palette of characteristics arises when nonuniform plane waves are considered in anisotropic and bianisotropic materials \c{ML_PRB}.

Lastly,  planewave solutions are used to construct surface waves guided by the planar interfaces of two dissimilar mediums. Accordingly, the 
 so-called ghost surface waves, including ghost Dyakonov waves \c{Nar_PRA} and ghost surface phonon polariton waves \c{JOSAB}, are simply  manifestations of nonuniform surface waves.

\vspace{10mm}
\noindent {\bf Acknowledgments.}
TGM acknowledges the support of EPSRC grant EP/S00033X/1.
AL thanks the Charles Godfrey Binder Endowment at the Pennsylvania State University for ongoing support of his research, with partial funding from US NSF grant number DMS-1619901.

\end{document}